\documentclass[epjST]{svjour}
\usepackage{graphicx}
\usepackage{amsmath}
\usepackage{amssymb}
\usepackage{mathrsfs}
\usepackage{float}
\usepackage[dvipsnames]{xcolor}
\usepackage{color}

\DeclareMathOperator*{\err}{err}

\begin{document}
\title{Sub-diffusive behavior in the Standard Map}
\author{Matheus S. Palmero\inst{1}\fnmsep\thanks{\email{palmero@usp.br}} \and Gabriel I. D\'iaz\inst{1} \and Iber\^e L. Caldas\inst{1} \and Igor. M. Sokolov\inst{2}}
\institute{Instituto de F\'isica, Universidade de S\~ao Paulo, S\~ao Paulo, SP, Brazil 
\and 
Institut für Physik, Humboldt-Universität zu Berlin, Berlin, Germany}
\abstract{In this work, we investigate the presence of sub-diffusive behavior in the Chirikov-Taylor Standard Map. We show that the stickiness phenomena, present in the mixed phase space of the map setup, can be characterized as a Continuous Time Random Walk model and connected to the theoretical background for anomalous diffusion. Additionally, we choose a variant of the Ulam method to numerically approximate the Perron-Frobenius operator for the map, allowing us to calculate the anomalous diffusion exponent via an eigenvalue problem, compared to the solution of the Fractional Diffusion Equation. The results here corroborate other findings in the literature of anomalous transport in Hamiltonian maps and can be suitable to describe transport properties of other dynamical systems.}


\maketitle

\section{Introduction}
\label{sec:intro}

Sub-diffusive processes are present in several areas of natural sciences \cite{Silvestri2009,Weiss2007,Weiss2004,Scher1975}. The fundamental understanding of these phenomena is necessary due to the amounting experimental evidence of sub-diffusive behavior in complex physical, chemical and biological systems \cite{Caspi2000,Saxton2006,Golding2006}. A process is determined to be sub-diffusive when the {\it mean squared displacement} does not grow linearly in time, presenting a growth exponent smaller than one. This means that the process is significantly slower than normal diffusion. Also, by definition, sub-diffusion is a type of anomalous diffusion, a matter with great scientific interest.

The evolution of systems described by non-integrable Hamiltonians often exhibits regularity and chaos. The result is a mixed phase space containing chaotic seas, invariant tori, and Kolmogorov-Arnold-Moser (KAM) islands \cite{Lichtenberg1992}. For strongly chaotic systems,  the diffusion of a chaotic orbit through its phase space is considered to be normal, with no anomalous effects \cite{Ott1979,Zaslavsky2007}. For mixed-phase spaces, however, a chaotic orbit evolved from initial conditions near a KAM island may exhibits very complicated dynamical behavior, influenced by its neighborhood and other structures that interfere with the orbits transport. In this scenario, the diffusion is often anomalous \cite{Venegeroles2008,Diaz2019}. 

Generally, the anomalous diffusion of orbits in the phase space of Hamiltonian systems is due to {\it stickiness} effect \cite{Zaslavsky2002,Altmann2013}. A chaotic orbit evolved from an initial condition set on a region surrounded by stability islands often experience {\it dynamical trappings}, spending a considerable amount of time in sticky regions around these islands and its cantori \cite{Zaslavsky1985,Balescu1997}. Applications of this trapping phenomenon can be found in many research areas as: fluid mechanics \cite{Solomon1993}, plasma physics \cite{delCastillo-Negrete2005}, celestial mechanics \cite{Contopoulos2010}, acoustics \cite{Altmann2009}, and biology \cite{Tel2005}. Additionally, studies on transport properties of chaotic orbits in Hamiltonian systems can be extended and applied to complex networks analysis \cite{Zou2016,Posadas-Castillo2014}.

In this work, we investigate the presence of sub-diffusive behavior in the Chirikov-Taylor Standard Map. First, we rely on the definitions of ergodicity and mixing to properly characterize the dynamic behavior of an ensemble of orbit evolved from an Unstable Period Orbit (UPO). The evolution started on the UPO guarantees stickiness influence in the chosen setup for the Standard Map. Based on the Mixing measure, we show that this dynamics can be described by a Continuous Time Random Walk (CTRW) model. In that sense, we chose to study the Fractional Diffusion Equation (FDE) to approach this problem. The solution of the FDE can be written in terms of the Mittag-Leffler function that depends explicitly on the anomalous diffusion exponent. Then, to be able to numerically calculate the anomalous diffusion exponent for the Standard Map, we choose a variant of the Ulam method. This method allows us to approximate the Perron-Frobenius operator for chaotic Hamiltonian maps. With that, it is possible to calculate the diffusion modes via an eigenvalue problem, and compare the computed eigenvalues to a fit provided by the Mittag-Leffler function, establishing a connection to the solution of the FDE. This comparison grants us an approximated value of the anomalous diffusion exponent that indicates a sub-diffusive behavior in the Standard Map, corroborating with other findings in the literature of stickiness and anomalous transport in Hamiltonian maps.

The paper is organized as follows. In Sec.\ \ref{sec:stdmap} we revise key aspects of the Standard Map. In Sec.\ \ref{sec:erg_mix} we briefly discuss ergodicity and import the definition of mixing to properly characterize the dynamical behavior of the selected setup for the Standard Map. In Sec.\ \ref{sec:anoma_diff} we review the CTRW model and discuss the analytical description of anomalous diffusion via the FDE. In Sec.\ \ref{sec:ulam_method} we present our selected numerical procedure to connect the dynamics in the Standard Map to anomalous diffusion, also referring to the Appendix where we further elucidate our numerical analysis. Finally, in Sec.\ \ref{sec:concluds} we conclude the chosen approach in this work.

\section{Standard Map}
\label{sec:stdmap}

The Standard Map \cite{Chirikov1969,Chirikov1979} can be used to describe the motion of a particle constrained to a movement on a ring while kicked periodically by an external field. 

The map derives from the Hamiltonian
\begin{eqnarray}
H\left(q,p,t\right) & = & \frac{p^2}{2mr^2}+K\cos\left(q\right)\sum_{n=-\infty}^{\infty}\delta\left(\frac{t}{T}-n\right)~, 
\label{eq:ham_stdmap}
\end{eqnarray} 
where $\delta$ is the Dirac delta function, $q$ is the angular coordinate and $p$ is its conjugate momentum. It is worth remarking that we consider the particle's mass $m$, the ring's radius $r$, and the period of kicks $T$ all equal to one for simplicity.

Although the Hamiltonian in Eq.\ (\ref{eq:ham_stdmap}) is sufficient to analyse the dynamics of our interest, it is possible to define a sympletic non-linear discrete map $T_{SM}$ to investigate the dynamics via extensive and long numerical simulations, considering that iterating the sympletic map is much faster than solving the equations of motion. 

\begin{figure}[h]
\includegraphics[scale=0.5]{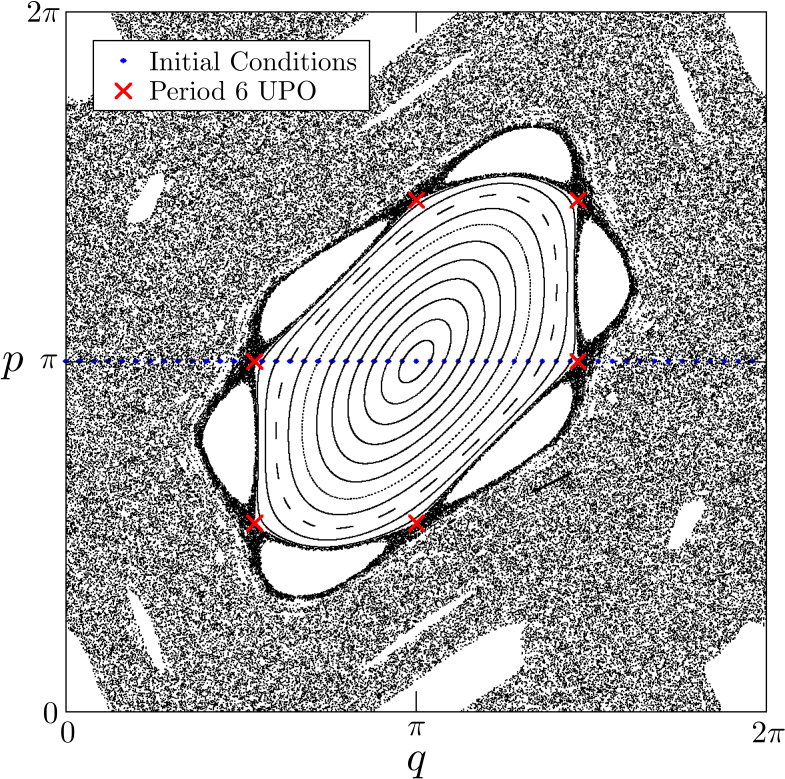}
\caption{Phase space of the Standard Map for $k=1.46$. In blue the chosen initial conditions iterated until $10^5$ and in red the period $6$ Unstable Periodic Orbit (UPO).}
\label{fig:phase_space}
\end{figure}

The mapping $T_{SM}\left(p_{n},q_{n}\right)=\left(p_{n+1},q_{n+1}\right)$ gives the position and momentum for the $(n+1)^{th}$ iteration by the following equations
\begin{equation}
T_{SM}:\left\{\begin{array}{ll}
p_{n+1} = p_n +k\sin(q_n) \mod(2\pi)\\
q_{n+1} = q_n + p_{n+1} + \pi \mod(2\pi)\\
\end{array}
\right.,
\label{eq:stdmap}
\end{equation}
where the parameter $k$ controls the intensity of the non-linearity and the added term $+\pi$, on the equation for $q_{n+1}$, is to centralize the main island on the drawn phase spaces. It is also important to mind that this is an area-preserving map since the determinant of its Jacobian matrix is equal to unity. The phase space for the selected control parameter $k=1.46$ is drawn in Fig.\ \ref{fig:phase_space}.

One can notice in Fig.\ \ref{fig:phase_space} that the region between the main stability island and resonant period-6 satellite islands is denser than the outer chaotic sea region. This is a classic evidence of stickiness. In this case, orbits started from this region experience successive traps, filling this region more densely at first. Then, after a sufficient number of iterations, the orbits, once trapped, are free to access other regions of the chaotic sea. 

In Fig.\ \ref{fig:phase_space} we are also drawing in red a period-6 UPO, associated with the period-6 satellite stable islands. The position of this UPO was calculated numerically with a floating-point double-precision via the method outlined in \cite{Ciro2018}. It is important to properly determine the position of this selected UPO because it will set our ensemble of initial conditions for the upcoming analysis.

\section{Ergodicity and Mixing}
\label{sec:erg_mix}
In physics, ergodicity is commonly defined by the equivalence of the ensemble and time averages of a relevant observable $y$, provided such averages formally exists: 

\begin{equation}
    \langle y\rangle_{ens}=\lim_{T\rightarrow\infty}\frac{1}{T}\int_0^{T}y(t^{\prime})dt^{\prime}~.
    \label{eq:ergodicity}
\end{equation}Since the right-hand side cannot depend on time, the process has to be stationary (so that the left-hand side is 
time-independent) and possess an invariant probability measure, over which the ensemble mean in the left-hand side 
is calculated. 

Experimentally\footnote{In our case, ``experimentally'' means in numerical simulations written in JULIA language.} this means that if Eq. (\ref{eq:ergodicity}) holds, averaging many single-time measurements over repeated experiments on the ensemble of similar systems, yields the same results as a time average over a single, but prolonged experimental run \cite{Meroz2015}. In experimental data, it is necessary to use a discrete-time variant of Eq. (\ref{eq:ergodicity}) given by 

\begin{equation}
    \langle y\rangle_{ens}=\lim_{N\rightarrow\infty}\frac{1}{N}\sum^{N-1}_{n=0}y_n~.
\end{equation}

Mixing, however, is the asymptotic independence of $y_n$ and $y_0$, as the step number $n$ goes to infinity. Here $y_n$ is our relevant observable and it behaves as random variable. Mixing is a stronger property than ergodicity \cite{Lasota1994} because all mixing processes are ergodic, but not vice versa. Proving mixing is enough to show ergodicity, and in some cases this may be an easier task.

In \cite{Garbaczewski1995,Janicki1993}, a 1D dynamical functional is defined by

\begin{equation}
    D_n=\langle e^{i(y_n-y_0)} \rangle~.
    \label{eq:dyn_func}
\end{equation}With that, the dynamical functional test \cite{Magdziarz2011} is based on the fact that $E_n = D_n - |\langle e^{i y_0} \rangle|^2$ vanishes when the random variables $y_n$ and $y_0$ are independent. Thus, a stationary process $y_n$ is said to mix if and only if

\begin{equation}
    \lim_{n\rightarrow \infty}E_n=0~.
    \label{eq:mixi}
\end{equation}

To show that the process is mixing, ensemble averages need to be calculated. Moreover, the process is ergodic if and only if

\begin{equation}
    \lim_{n\rightarrow \infty}\frac{1}{n}\sum_{j=0}^{n-1}E_j=0~.
    \label{eq:erg_mix}
\end{equation}

Note that the tests are quite restrictive with respect to the amount of data, which has to be sufficiently large. 

\begin{figure}[h]
\includegraphics[scale=0.5]{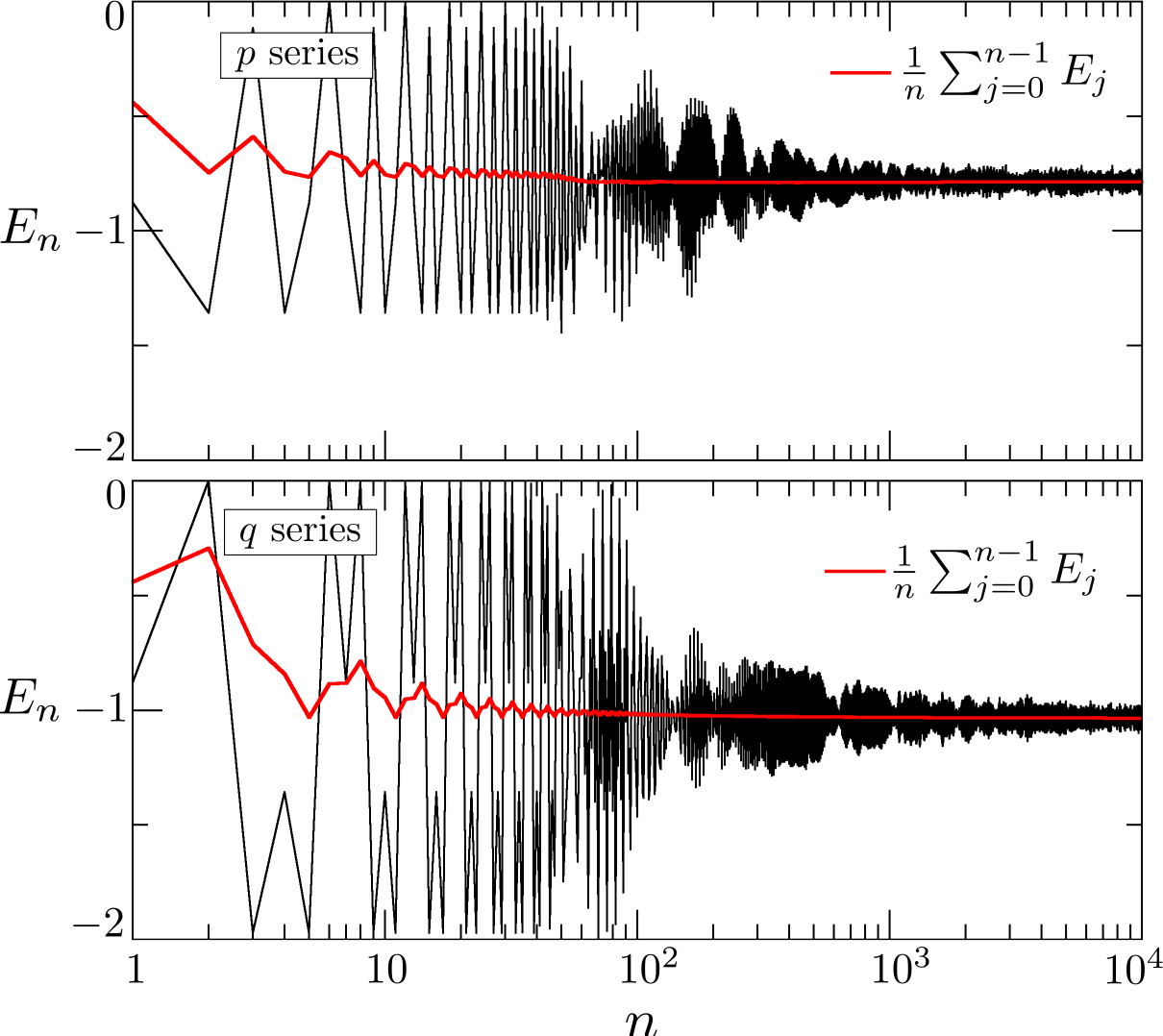}
\caption{Mixing measure for the $p$ and $q$ series, considering an evolution started at the period-6 UPO for the selected value of the control parameter $k=1.46$ of the Standard Map.}
\label{fig:erg_mix_series}
\end{figure}

For the coordinates $q$ and $p$ of the Standard Map, we make use of the mixing definition considering $y_n=p_{n+\Delta n}-p_n$ and $y_n=q_{n+\Delta n}-q_n$, as the increment process for each coordinate. We evolved an ensemble of $10^3$ initial conditions in a small square of size $10^{-4}$, centered at the position of the period-6 UPO calculated before and, with that we analyse the time-series of $q$ and $p$. We show in Fig.\ \ref{fig:erg_mix_series} the asymptotic behavior of $E_n$ for $p$ and $q$ series.

From Fig.\ \ref{fig:erg_mix_series} it is clear that either $p$ or $q$ does not present ergodic behavior, since Eqs.\ (\ref{eq:mixi},~\ref{eq:erg_mix}) do not hold. This is expected because the selected setup for the Standard Map has a mixed phase space, evidenced in Fig.\ \ref{fig:phase_space}. A mixed phase space contains regions of stability along with a chaotic sea, however, due to this mix, other relevant structures that influence the transport of orbits are also present, this is the case of invariant manifolds associated with the dynamics. It is also important to note that we chose the ensemble centered at the UPO, guaranteeing a dynamics influenced not only by the stickiness but also by the unstable and stable branches of the manifold associated with this UPO. An ensemble evolved from other regions of the chaotic sea may present a dynamic behavior that agrees with Eqs.\ (\ref{eq:mixi},~\ref{eq:erg_mix}).

Furthermore, the results of the mixing measure for the dynamics of the Standard Map, shown in Fig.\ \ref{fig:erg_mix_series}, allow a different interpretation for this dynamics. According to the decision tree proposed in \cite{Meroz2015}, since the trajectories of our interest exhibit a non-ergodic dynamical behavior, they can be modeled by the Continuous Time Random Walk (CTRW) description. We explain the CTRW model in detail and connect it to a theoretical description of anomalous diffusion in the next section.

\section{Continuous Time Random Walk and Anomalous Diffusion}
\label{sec:anoma_diff}

The CTRW model was introduced by Scher and Montroll when they investigated the anomalous transport properties of charge carriers in amorphous materials \cite{Scher1975}. The model is given by successive traps of a random walk, in which each step is characterized by two independent stochastic processes, the waiting time and the displacement in space. The combination of these two factors yields an anomalous diffusion process. It is important to note that, because of the basic nature of CTRW, it is a well-suited candidate to describe stickiness phenomena in Hamiltonian systems.

Since CTRW is an example of anomalous diffusion, we studied the Fractional Diffusion Equation (FDE), one of the main theoretical descriptions of anomalous diffusion. Further discussions and examples of CTRW simulations are presented in the Appendix. The FDE is given by

\begin{eqnarray}
\frac{\partial^{\alpha}\rho\left(x,t\right)}{\partial t^{\alpha}} & = & D\frac{\partial^{2}\rho\left(x,t\right)}{\partial x^{2}}~,
\label{eq:FDE}
\end{eqnarray}note the presence of the operator $\frac{\partial^{\alpha}}{\partial t^{\alpha}}$, (Caputo fractional derivative with $0<\alpha\leq1$) \cite{Haubold2011}. The parameter $\alpha$ represents the anomalous diffusion exponent, and $D$ the
diffusion coefficient. 

In a closed interval $0\leq x\leq 2\pi$ with periodic boundary conditions, the solution of Eq. (\ref{eq:FDE}) is given by the following superposition \cite{Haubold2011}

\begin{equation}
   \text{\small$ 
   \rho\left(x,t\right) =  a_0+\sum_{m=1}^{\infty}E_{\alpha}\left(-D m^{2}t^{\alpha}\right)\left(a_{m}\cos\left(mx\right)+b_{m}\sin\left(mx\right)\right)~,$}
   \label{eq:supsolFDE}
\end{equation}where the first term in the right-hand side can be set to $a_0=\frac{1}{2\pi}$ due to normalization of
$\rho\left(x,t\right)$. The terms $a_{m}$ and $b_{m}$ are related to the initial condition $\rho\left(x,0\right)$ and $E_{\alpha}\left(z\right)$ is the Mittag-Leffler function defined by

\begin{eqnarray}
E_{\alpha}\left(z\right) & = & \sum_{k=0}^{\infty}\frac{z^{k}}{\Gamma\left(1+\alpha k\right)}~.
\label{eq:MittagL}
\end{eqnarray}

Here $\alpha, z \in \mathbb{C}$, $\Re\left(\alpha\right)>0$, $\mathbb{C}$ being the set of complex numbers and  $\Gamma\left(z\right)$ is the Gamma function \cite{Haubold2011}.
 
Regarding the solution presented in Eq.\ (\ref{eq:supsolFDE}) and the Mittag-Leffler function defined in Eq.\ (\ref{eq:MittagL}) the following observations can be made:

\begin{enumerate}
\item The density $\rho(x,t)$ approaches to a limit $\rho\sim\frac{1}{2\pi}$; 

\item Besides of the first term $a_0$ of Eq. (\ref{eq:supsolFDE}), all other terms of the summation have zero integral in space, e.g. $\int_{0}^{2\pi}\cos\left(mx\right)dx=0$, etc; 

\item The Mittag-Leffler function equals to the exponential when $\alpha=1$; 

\item For values of $0<\alpha<1$ the decay given by $E_{\alpha}\left(-t^{\alpha}\right)$ is slower than exponential but the limit $\lim_{t\rightarrow\infty}E_{\alpha}\left(-t^{\alpha}\right)=0$ still holds.
\end{enumerate}

Furthermore, one can define

\begin{eqnarray}
\text{\small$ 
\hat{M} \left(x,y,t\right) = \frac{1}{2\pi}+\frac{1}{\pi}\sum_{m=1}^{\infty}E_{\alpha}\left(-D m^{2}t^{\alpha}\right)\left[\cos\left(my\right)\cos\left(mx\right)+\sin\left(my\right)\sin\left(mx\right)\right]~,$}
\label{eq:Op_FDE}
\end{eqnarray} and rewrite Eq. (\ref{eq:supsolFDE}) as following

\begin{eqnarray}
\rho\left(x,t\right) & = & \int_0^{2\pi} dy ~\hat{M} \left(x,y,t\right) \rho\left(y,0\right)~.
\label{eq:TransOp_flow}
\end{eqnarray} 

Then, Eq.\ (\ref{eq:TransOp_flow}) defines a Perron-Frobenius-like transfer operator by virtue of the properties of the solutions $\rho\left(x,t\right)$ in Eq. (\ref{eq:FDE}). In one hand, the eigenfunctions of the operator, given by $\left[1,~\cos\left(mx\right),~\sin\left(mx\right)\right]$, $m\in\mathbb{Z}^{+}$ are time-independent. On the other, the eigenvalues, given by $\left[1,~E_{\alpha}\left(-Dm^{2}t^{\alpha}\right)\right]$, carry the time dependence. 

One can observe that changing the right-hand side of Eq. (\ref{eq:FDE}) for other linear position-dependent operator, only changes the eigenfunctions and not the eigenvalues. Also, the presence of $m$ in the argument of the Mittag-Leffler function  makes the influence of the eigenfunctions with higher $m$ go to zero fast. Then, it is fair to assume that, in the asymptotic time $t\rightarrow\infty$, the transient behavior is governed by the first eigenfunction.   

A discrete scenario of Eq.\ (\ref{eq:TransOp_flow}) can be used to analyze the dynamics governed by mappings. The discrete version of Eq.\ (\ref{eq:TransOp_flow}) is given by

\begin{eqnarray}
\rho\left(i,n\right) & = & \sum_{j} ~\hat{S} \left(i,j,n\right) \rho\left(j,0\right)~,
\label{eq:TransOp_disc}
\end{eqnarray} where $i,j$ are {\it space-like} indexes and $n$ a {\it time-like}. Assuming that $[\lambda_l,\psi_l]$ are the eigenpairs associated to $\hat{S}\left(i,j,n\right)$ and, since this matrix is a Perron-Frobenius-like operator, the following remarks are noted:

\begin{enumerate}
\item The index $l$ numerates the eigenpairs $[\lambda_l,\psi_l]$;
\item All eigenvalues of $\hat{S}\left(i,j,n\right)$  have the constrain $\left|\lambda_{l}\left(n\right)\right| \leq 1$;
\item There is at least one eigenvector $\psi_{0}\left(i\right)$ with eigenvalue $\lambda_{0}=1$ and $\sum_{i}\psi_{0}\left(i\right)=1$;
\item Eigenvectors $\psi_{l}$ associated to $|\lambda_l|<1$, have  $\sum_{i}\psi_{l}\left(i\right)=0$. Otherwise the probability would not be conserved. 
\end{enumerate}

For a Markovian process ${\footnotesize\hat{S}\left(i,j,n\right)=\left(\hat{S}\left(i,j,1\right)\right)^{n}}$, and the eigenvalues of $\hat{S}\left(i,j,n\right)$ are given by $\lambda\left(n\right)=\lambda^{n}$, where $\lambda$ represents the eigenvalues of $\hat{S}\left(i,j,1\right)$. In the case that there is just one eigenvalue with $\lambda_{0}=1$, the time dependence of the probability distribution is given by

\begin{eqnarray}
\rho\left(i,n\right) & = & \psi_{0}\left(i\right)+\sum_{l}\exp\left(n\log\left(\lambda_{l}\right)\right)a_{l}\left(0\right)\psi_{l}\left(i\right)~,
\label{eq:normaldiffsol}
\end{eqnarray} one can note that, in this case, the solution corresponds to the superposition solution shown in Eq.\ (\ref{eq:supsolFDE}) with $\alpha=1$.

For a non-Markovian process in general ${\footnotesize\hat{S}\left(i,j,n\right)\neq\left(\hat{S}\left(i,j,1\right)\right)^{n}}$ and the solution given by Eq. (\ref{eq:normaldiffsol}) is not longer valid. Nevertheless, the following behavior shown in Eq. (\ref{eq:eigenvalue_Mittag}) can be valid if the diffusion process is believed to come from a CTRW-like model
\begin{eqnarray}
\lambda_l\left(n\right) & = & E_{\alpha_l}\left(-d_l n^{\alpha_l}\right)~,
\label{eq:eigenvalue_Mittag}
\end{eqnarray}where we define $\alpha_l$ and $d_l$ as, respectively, the anomalous diffusion exponent and coefficient related to the $l^{th}$eigenpair, see the Appendix for further discussion and examples.

Our approach in this work is focused in the eigenpair with the highest eigenvalue smaller than one $[\lambda_1,\psi_1]$, since, as stated before, it is the one that governs the transient dynamics of $\rho$ in the asymptotic limit  $n\rightarrow\infty$.

Finally, is important to note from Eq. (\ref{eq:eigenvalue_Mittag}), that it is possible to assess the anomalous diffusion exponent $\alpha_1$, associated to the assumed most relevant eigenpairs $[\lambda_1,\psi_1]$. Hence, the problem now is to calculate the Perron-Frobenius-like operator for the selected setup of the Standard Map to characterize the most relevant dynamic behavior via $\alpha_1$. In the next section we discuss a numerical approach to this problem.

\section{Numerical approach - Ulam method}
\label{sec:ulam_method}

\begin{figure}[b]
\includegraphics[scale=0.28]{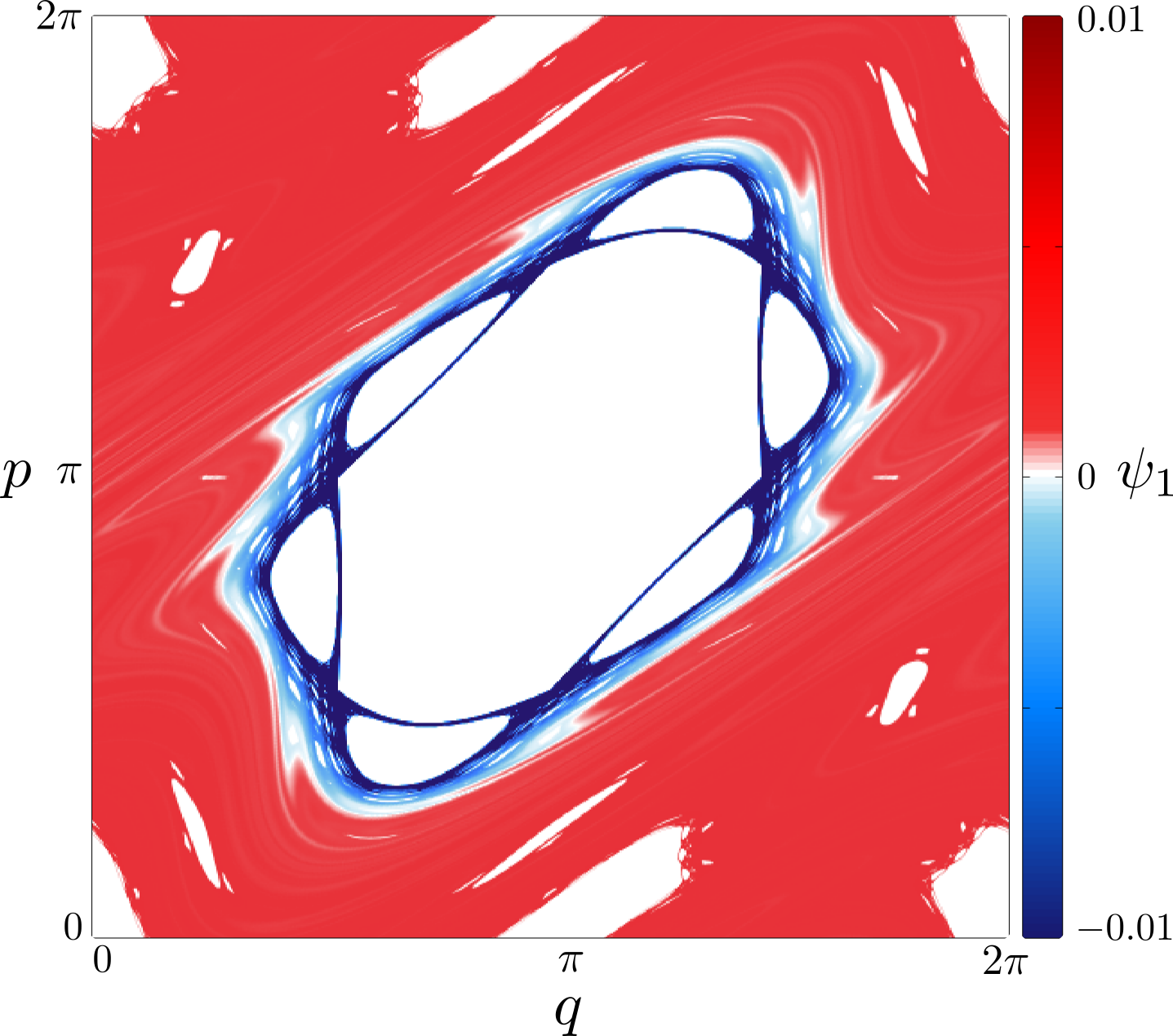}
\caption{First diffusion mode, given by the first eigenvector $\psi_1$ of $\hat{S}\left(n^{\prime}\right)$, on the phase space of the standard map for the selected control parameter $k=1.46$. We considered the space divided by a grid of $512\times512$ cells.}
\label{fig:diff_mode}
\end{figure}

To be able to simulate the discrete version of the transfer operator equation discussed before, considering the dynamics of the selected setup for the Standard Map, we make use of a variant of the Ulam method proposed by Frahm and Shepelyansky in \cite{Frahm2010,Frahm2013}. 

The chosen numerical method is used to approximate the Perron-Frobenius operator for chaotic Hamiltonian maps. It is based on following the evolution of one chaotic orbit through a discretized phase space. First, we need to divide the phase space into a regular grid and, at each iteration, ask, in which cell $j$ the orbit is, and at which cell $i$ it jumps in the next iteration. After many iterations, we calculated how many times an orbit in cell $j$ jumped to cell $i$. Then, it is defined the matrix elements $S_{ij}$. Considering a normalization condition $\sum_{i}S_{ij}=1$, that give us the probability to jump from $j$ to $i$ after one iteration ($n^{\prime}=1$). Additionally, if we consider the cell $i$ that we arrive from cell $j$ after $n^{\prime}$ iterations we have the matrix element $S_{n^{\prime},ij}$. Due to the normalization condition, the matrices $\hat{S}\left(n^{\prime}\right)$ with elements $S_{n^{\prime},ij}$, are Perron-Frobenius-like operators, which have at least one invariant probability vector. It is important to note that we differentiate $n$ from $n^\prime$ as, respectively, the number of iterations for the orbit's evolution and the iterations skipped to calculate the Perron-Frobenius-like operator.

The eigenvectors, in this case, are called diffusion modes and are defined by the eigenvalue problem

\begin{eqnarray}
\sum_{j}S_{ij}\psi_{l}\left(j\right) & = & \lambda_{l}\psi_{l}\left(i\right)~,
\end{eqnarray}
where $\lambda_{l}$ is the eigenvalue and $\psi_{l}\left(i\right)$ is the value of the eigenvector $\psi_{l}$ at the cell $i$. Note that $\psi_{l}$ may have different numerical values for different cells in the correspondent region of the discrete space. We call the first diffusion mode the $\psi_{l}$ that has the largest $\left\Vert \lambda_{l}\right\Vert \neq1$. 

In discrete spaces is often necessary to set a fine grid that guarantees enough resolution to distinguish small structures. However, a fine grid combined with sufficient $n^{\prime}$ iterations to calculate the matrices, demand powerful computational resources. To surpass this issue, we make use of the Arnoldi method described in \cite{Frahm2010,Frahm2013}. With that, we are able to approximate the eigenpairs $\left[\lambda_l,\psi_l \right]$. 

In Fig.\ \ref{fig:diff_mode} we show the numerical result for the first diffusion mode $\psi_1$ through the phase space of the Standard Map, considering only one trajectory started from the selected period-6 UPO, but evolved until $10^{10}$ iterations. For this first result we considered $n^{\prime}=1$, since $\psi_1$ does not change enough to modify this analysis. This matter will be addressed further in this section.

\begin{figure}[b]
\includegraphics[scale=0.42]{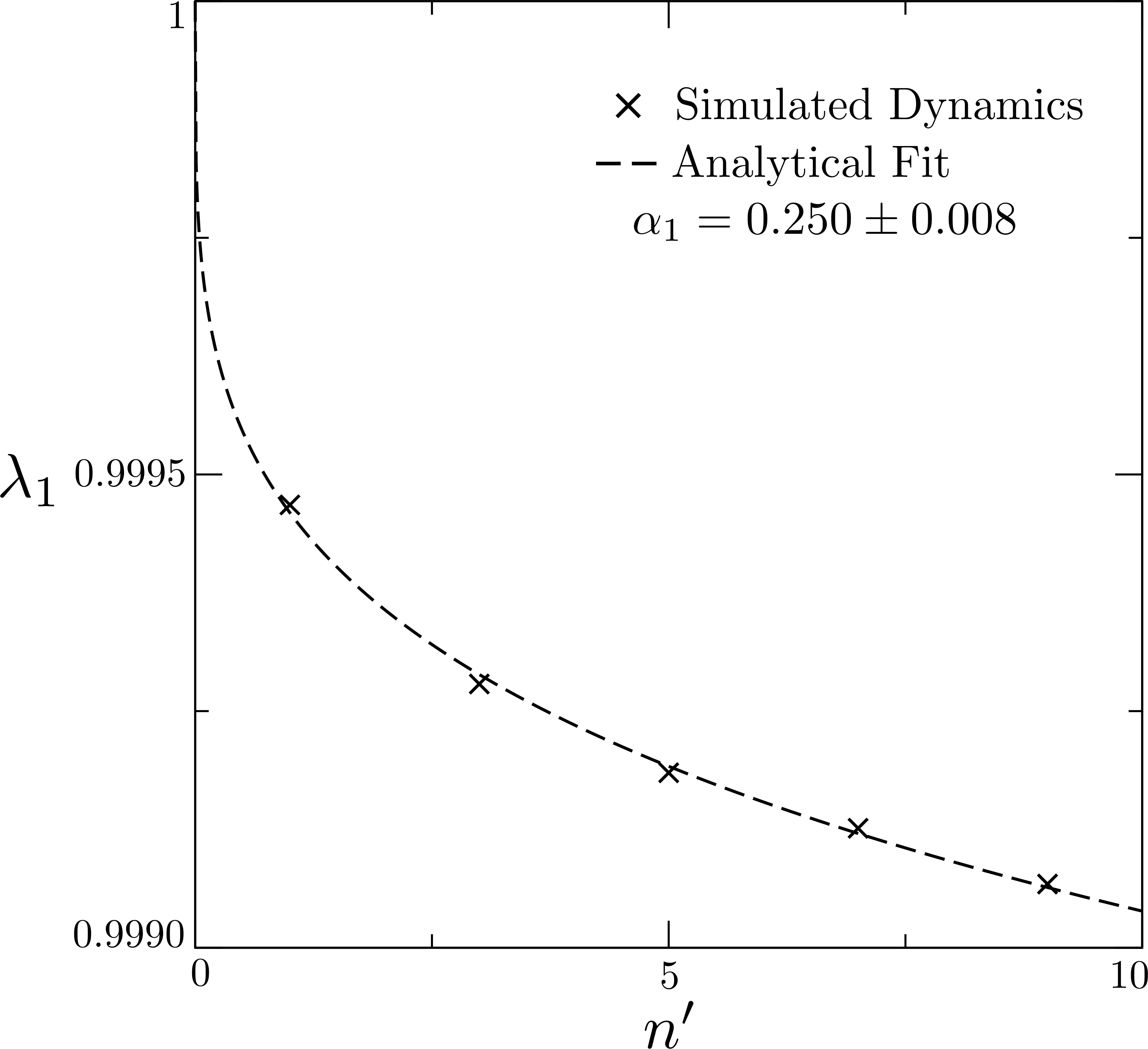}
\caption{Eigenvalue $\lambda_{1}$ of matrix $\hat{S}\left(n^{\prime}=1\right)$ after skipping $n^{\prime}$ iterations for the Standard map with $k=1.46$. The analytical fitting was provided by a numerical representation for the Mittag-Leffler function.}
\label{fig:fit_std_map}
\end{figure}

It is clear from the color map in Fig.\ \ref{fig:diff_mode} that the phase space is divided in two distinct regions. This division is properly characterized by the change of sings of the first diffusion mode $\psi_1$. Negative values of $\psi_1$ are attained only at the region around the main island and the other resonant ones. Yet, positive values of $\psi_1$ depicts the chaotic sea.
The regions with $\psi_1=0$ are also interesting, but will be investigated in another study.   

One can notice that the region of negative $\psi_1$ is close related to the region of stickiness influence, as depicted early in Fig.\ \ref{fig:phase_space}. This establishes a first connection between the dynamics affected by the stickiness and its diffusion mode. To further investigate this special region, we analyse the behavior of the first eigenvalue $\lambda_1$, associated with the first diffusion mode, as function of the first nine iterations skipped to calculate the Perron-Frobenius-like operator. Moreover, based on the analytical description made in last section, we propose fitting the numerical values of $\lambda_1$ with a numerical representation \cite{Gorenflo2002,Valerio2014} for the Mittag-Leffler function at Eq.\ (\ref{eq:MittagL}). This result is shown in Fig.\ \ref{fig:fit_std_map}.

The connection between the computed eigenvalues $\lambda_1=\lambda_1(n^{\prime})$ and the fit provided by the Mittag-Leffler function is further addressed and tested on the Appendix. There we simulate the CTRW model for different values of the anomalous diffusion coefficient $\alpha$ and show the reliability of this procedure.

\begin{figure}[b]
\includegraphics[scale=0.42]{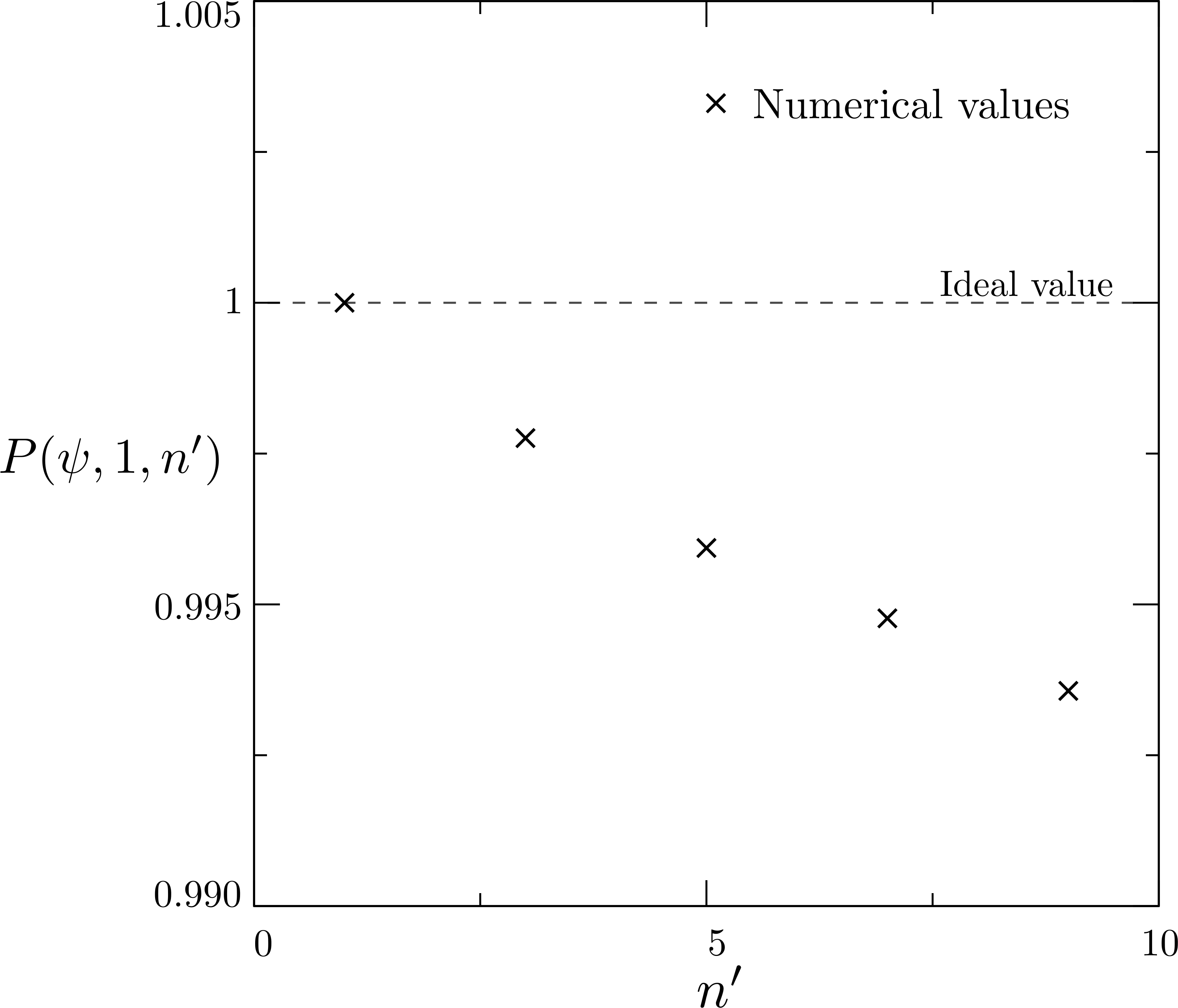}
\caption{Dependence of the error function defined in Eq.\ (\ref{eq:error_fit}) with the number of iterations skipped, i. e. the dot product between the diffusion mode after one iteration and the diffusion mode after $n^{\prime}$ iterations. The dashed line at $\err(\psi,1,n^{\prime})=1$ indicates the ideal value.}
\label{fig:error_fit_std_map}
\end{figure}

The result shown in Fig.\ \ref{fig:fit_std_map} provides a fitted value of $\alpha_1=0.250\pm0.008$ that implies a dependence slower than exponential for $\lambda_{1}(n^{\prime})$. This corroborates our initial assumption that the dynamic behavior of this setup for the Standard Map is indeed related to anomalous diffusion, specifically a sub-diffusive process with $\alpha\approx0.25$.

It is also important to mention that, when studying the dependence of $\lambda_{1}$ with the matrix iteration $n^{\prime}$, we must check if we are still considering the same diffusion mode, or, in each iteration, it is changing due numerical errors. To control this deviance from unity, we apply the normalization conditions $\sum_{i}\psi_{1}\left(i,t=1\right)\psi_{1}\left(i,t=1\right)=1$, $\sum_{i}\psi_{1}\left(i,t=n^{\prime}\right)\psi_{1}\left(i,t=n^{\prime}\right)=1$
and compute the value of the projection
\begin{equation}
P(\psi,1,n^{\prime})=\left\Vert\psi_{1}\left(t=1\right)\cdot\psi_{1}\left(t=n^{\prime}\right)\right\Vert=\left\Vert \sum_{i}\psi_{1}\left(i,t=1\right)\psi_{1}\left(i,t=n^{\prime}\right)\right\Vert~,
\label{eq:error_fit}
\end{equation}
i.e., the dot product between the diffusion mode after one iteration and the diffusion mode after $n^{\prime}$ iterations. 

In Fig.\ \ref{fig:error_fit_std_map} we show the computed projection $P(\psi,1,n^{\prime})$ for the first nine iterations skipped to calculate the Perron-Frobenius-like operator. Since the error does not diverge from its ideal value of unity, considering only the first decade, it is fair to assume that our analysis of $\alpha\approx0.25$ still holds. However, for larger $n^{\prime}$, we are not dealing with the same diffusion mode, and the calculation of $\alpha$ becomes meaningless. 

It is important to note that with the described numerical procedure is possible to calculate the most relevant diffusion mode $\psi_1$, and associated with it, the anomalous diffusion exponent $\alpha_1\approx\alpha$ that characterize the predominant dynamical behavior for the time evolution of interest. In the case of the selected setup for the Standard Map, the anomalous sub-diffusive behavior is confirmed by $\alpha\approx0.25$. Furthermore, one can adapt the procedure to investigate other dynamical systems, particularly to study sub-diffusive transport in complex networks.

\section{Conclusions}
\label{sec:concluds}

In this work, we present a procedure, based on different descriptions found in the literature of anomalous diffusion, to show that the evolution of trajectories under stickiness influence in the Chirikov-Taylor Standard Map can indeed be described as a sub-diffusive process. 

Initially, we showed that for the chosen control parameter $k=1.46$, the Standard Map presents a non-ergodic dynamical behavior, as expected for a mixed phase space. Then, we assumed that the trajectories in this setup can be described by the Continuous Time Random Walk model. Since CTRW is a classic example of anomalous diffusion, we studied some consequences of the Fractional Diffusion Equation and how to connect it to a numerical method that approximates the Perron-Frobenius operator for the map. With that, we established a relation between the eigenvalues of the Perron-Frobenius operator and the solution of FDE, providing an approximated value for the anomalous diffusion exponent $\alpha$ for the selected setup of the Standard Map. The value of $\alpha\approx0.25$ shows that the evolution of trajectories in this scenario is indeed associated with anomalous diffusion, distinctively a sub-diffusive behavior. Moreover, we connected the stickiness in the Standard Map with a robust framework to describe this effect as a sub-diffusive anomalous transport.

It is important to note that the procedure described here is readily applicable to other Hamiltonian systems with a variety of applications and also suitable to investigate transport properties on complex networks.   

\section*{Acknowledgments}
This study was financed in part by the Coordena\c c\~ao de Aperfei\c coamento de Pessoal de N\'ivel Superior - Brazil
(CAPES) - Finance Code 001, S\~ao Paulo Research
Foundation (FAPESP) - Brazil, under Grants No. 2018/03211-6 and 2018/03000-5, and IRTG 1740 financed by Deutsche Forschungsgemeinschaf (DFG).

\section*{Appendix: Stochastic representation and RW/CTRW Simulations}
\label{app:A}

This Appendix is devoted to verify and support our analysis made for the selected setup of the Standard Map, based on the connection between the eigenvalue of the Perron-Frobenius transfer operator and the Mittag-Leffler function.

We address here how to simulate a Random Walk (RW) and a Continuous Time Random Walk (CTRW) using the It\^o stochastic representation \cite{Magdziarz2007}. 

A classical CTRW is a process subordinated to a simple RW, in which steps (jumps) follow at random instants of time. The waiting time $t_i$ for the next step follows the probability distribution with the known probability density $\psi(t_i)$. The number of steps performed up to time $t$, $\nu(t)$ is the {\it operational time}, or the subordinator of the corresponding subordination scheme. The {\it clock time} $t$ of the $\nu$-th step is then the following sum
\begin{equation}
   t(\nu) = \sum_{i=1}^\nu t_i~, 
   \label{eq:time_sum}
\end{equation}which, for fat-tailed $\psi(t_i)\sim t^{-1-\alpha}$
with $0< \alpha< 1$ tends in distribution to a one-sided Levy law. The value of $\nu$ as a function of $t$ is then $\nu(t) = \mathrm{inf}\{ \nu, t(\nu)>t \}$ \cite{Sokolov2005}. This gives us the prescription for stochastic simulations of the CTRW. 

For long times the variable $\nu$ can be taken continuous; the corresponding continuous limit for the Probability Density Function $\rho(x,t)$ is 
then given by the Fractional Diffusion Equation showed in Eq.\ (\ref{eq:FDE}).

In the displacement $X(t)$ then follows from a couple of stochastic differential equations \cite{Fogedby1994}
\begin{eqnarray}
dX(\nu) &=& \sqrt{2D}dB(nu), \\
dt(\nu) &=& \tau d L_\alpha(\nu)
\label{eq:displace}
\end{eqnarray}with $B(\nu)$ being a standard Brownian motoin, $L_\alpha(\nu)$ a one-sided L\'evy flight, and $D$ and $\tau$ 
the appropriate constants. From this representation Eq.\ (\ref{eq:FDE}) follows as well. 

\begin{figure}[h]
\includegraphics[scale=0.45]{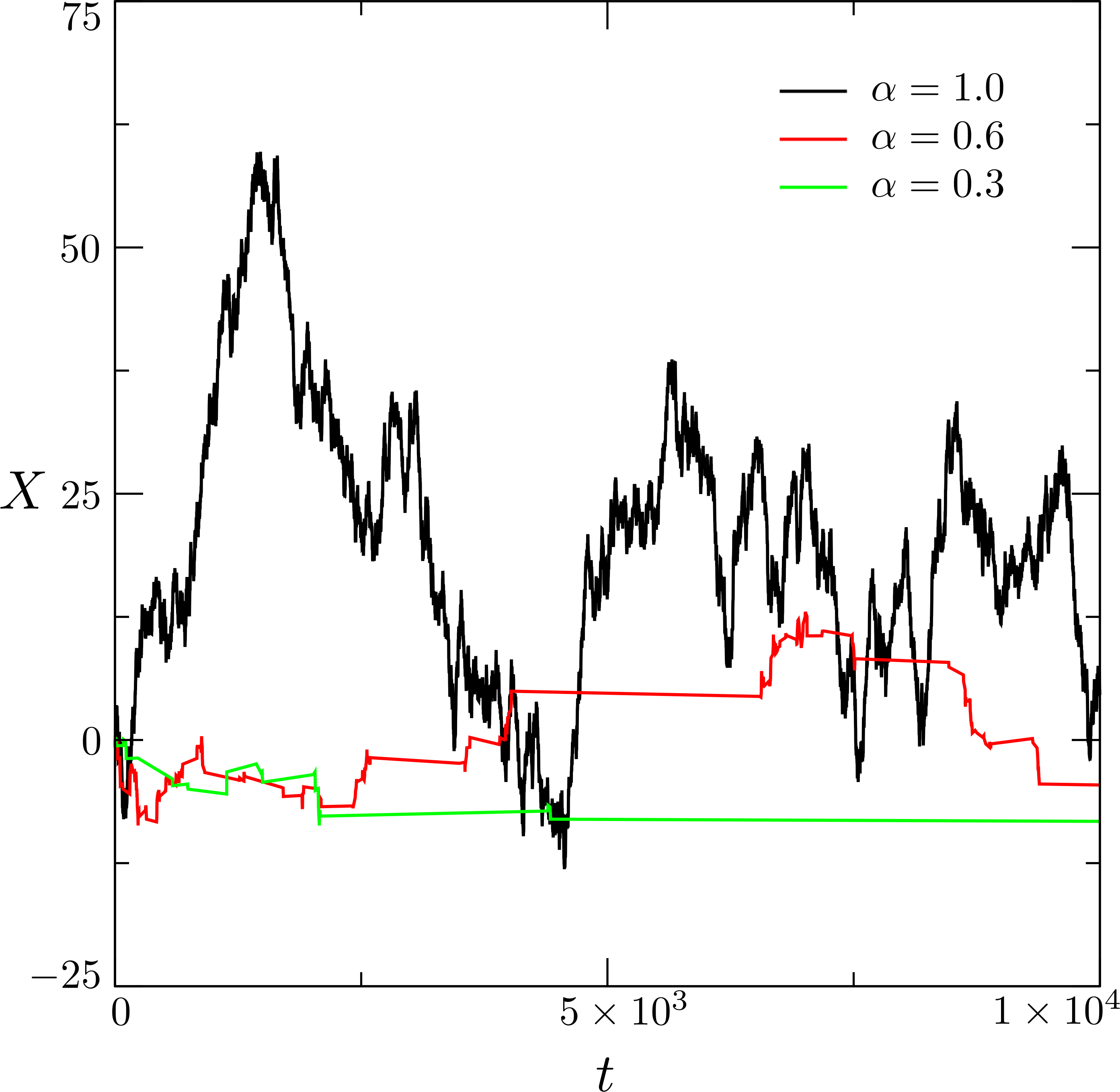}
\caption{Examples of trajectories for a RW process ($\alpha=1)$ in black. Also, for two distinct CRTW processes considering a relatively not so long trap-time ($\alpha=0.6$) in red, and a long trap-time ($\alpha=0.3$) in green. The time-axis was chosen to be linear to properly illustrate the different trap-times.}
\label{fig:ctrw_examples}
\end{figure}

With that, we followed the numerical method proposed by Magdziarz and Weron in \cite{Magdziarz2011,Magdziarz2007} to simulate the dynamics of a RW process, setting $\alpha=1$ and CTRW processes, setting $\alpha=0.6$ (relatively not so long trap-time) and $\alpha=0.3$ (relatively long trap-time). In Fig.\ \ref{fig:ctrw_examples} we show examples of simulated trajectories of these three different dynamic scenarios.

The discussed Ulam method, shown in Sec.\ \ref{sec:ulam_method}, used for approximate the Perron-Frobenius-like operator to approach the anomalous diffusion exponent $\alpha_1$ for the selected setup of the Standard Map, can also be applied to an ensemble of particles whose movement is governed by RW. However, it is not an appropriate method to analyze cases where $\alpha\neq1$. In that case of anomalous diffusion, it is necessary to apply a kernel density estimator to have a good approach to the probability density from a histogram \cite{Magdziarz2007}. To avoid the expensive use of a kernel density estimator, we choose to use a different method called the Extended Dynamic Mode Decomposition (EDMD) \cite{Willams2015,Klus2016}.

\begin{figure}[b]
\includegraphics[scale=0.45]{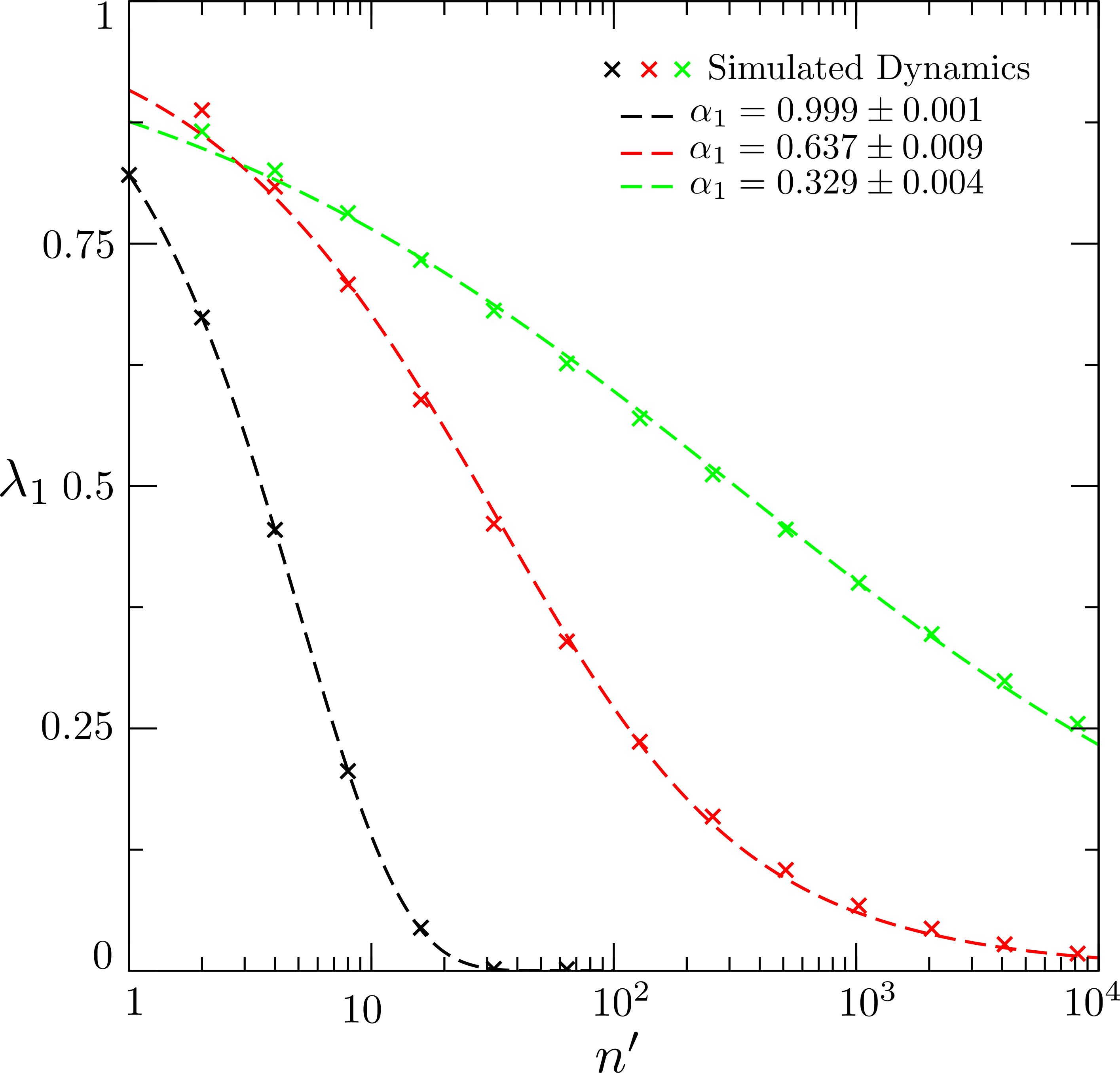}
\caption{Behavior of the eigenvalue $\lambda_{1}$ as function of the log of the number of iterations $n^{\prime}$ to calculate the correspondingly matrices, considering the dynamics of a RW and the two CTRW examples. The points mark the simulated dynamics in each case and the dashed line is the fitting provided by a numerical representation of the Mittag-Leffler function.}
\label{fig:ctrw_fitting}
\end{figure}

The EDMD in \cite{Klus2016} is method to approximate the Perron-Frobenius operator, using the fact this operator and the Koopman operator\footnote{The Koopman operator maps functions of state space to functions of state space.} are adjoint to each other. The EDMD method focus on a dictionary of observables $D=\left[\phi_1(x),\phi_2(x),\ldots,\phi_k(x)\right]$, functions of state space, and how they change along the trajectory. With this information, and the definition of the Koopman operator is possible to approximate the eigenfunctions and eigenvalues both of the Koopman operator as of the Perron-Frobenius operator. Much of the method relies in an educated guess for the dictionary of observables, that must be rich enough to approximate the Koopman operator eigenfunctions, see \cite{Willams2015,Klus2016} for further details.

We consider the region $0\leq x\leq2\pi$ with periodic boundary conditions and choose the observables
\begin{equation}
   \text{\small$D= \left[1,\cos\left(x\right),\sin\left(x\right),\cos\left(2x\right),\sin\left(2x\right),\ldots,\cos\left(5x\right),\sin\left(5x\right)\right]$}~,
   \label{eq:observ}
\end{equation}as the dictionary for the EDMD method. 

The behavior of the eigenvalue $\lambda_{1}$ as function of the skipped iterations $n^{\prime}$ is shown in Fig.\ \ref{fig:ctrw_fitting} for the same simulations depicted in Fig.\ \ref{fig:ctrw_examples}. Considering its relation to the solution of the FDE, as explained in Sec.\ \ref{sec:anoma_diff}, we selected the Mittag-Leffler function, given by 

\begin{eqnarray}
E_{\alpha}\left(z\right) & = & \sum_{k=0}^{\infty}\frac{z^{k}}{\Gamma\left(1+\alpha k\right)}~,
\label{eq:MittagL_}
\end{eqnarray}to fit these three different dynamic scenarios for three different values of $\alpha$.

The Mittag-Leffler is the suitable function because of the assumed behavior

\begin{eqnarray}
\lambda_l\left(n\right) & = & E_{\alpha_l}\left(-d_l n^{\alpha_l}\right)~,
\label{eq:eigenvalue_Mittag_}
\end{eqnarray}where it was defined $\alpha_l$ and $d_l$ as, respectively, the anomalous diffusion exponent and coefficient related to the $l^{th}$eigenpair of the Perron-Frobenius-like operator. 

Is important to mind that in these particular cases, the anomalous diffusion exponent $\alpha_l$ related to the $l^{th}$eigenpair, is the same as $\alpha_1$. Then, the anomalous diffusion exponent $\alpha$, that characterize the anomalous dynamics, is $\alpha\approx\alpha_1=\alpha_l$. 

Finally, one can observe that the values of the fitted functions are close to the actual values set for its simulations. Specially, for the CTRW model, the given values of $\alpha_1=0.637\pm0.009$ and $\alpha_1=0.329\pm0.004$ imply a dependence slower than exponential, as expected from the theory.

\end{document}